\documentclass[12pt]{article}
\usepackage{graphicx,epsfig,epsf}

\usepackage{graphicx}
\usepackage{scicite}
\usepackage{times}

\def\ltsima{$\; \buildrel < \over \sim \;$}
\def\simlt{\lower.5ex\hbox{\ltsima}} 
\def\gtsima{$\; \buildrel > \over \sim \;$}
\def\simgt{\lower.5ex\hbox{\gtsima}} 

\def\deg{\hbox{$^\circ$}}
\def\etal{{\it et al.}}

\def\aap{{\it Astron.~Astrophys.}}
\def\aip{{\it AIP Conf.~Ser.}}
\def\aj{{\it Astron.~J.}}
\def\apj{{\it Astrophys.~J.}}
\def\apjl{{\it Astrophys.~J.~Lett.}}
\def\apjs{{\it Astrophys.~J.~Suppl.~Ser.}}
\def\ajp{{\it Aust.~J.~Phys.}}
\def\araa{{\it Annu.~Rev.~Astron.~Astrophys.}}
\def\ijmpd{{\it Int. J. Mod. Phys. D}}
\def\mnras{{\it Mon.~Not.~R.~Astron.~Soc.}}
\def\njp{{\it New J.~Phys.}}  
\def\nat{{\it Nature}}

\newenvironment{sciabstract}{ \begin{quote} \bf} {\end{quote}}

\newcounter{lastnote}

\title{$Fermi$ Gamma-ray Imaging of a Radio Galaxy}

\author{The Fermi-LAT Collaboration\footnote{All authors with their 
affiliations appear at the end of this paper.}}

\date{}

\begin{document}

\baselineskip24pt

\maketitle

\begin{sciabstract} The $Fermi$ Gamma-ray Space Telescope has detected 
the $\gamma$-ray glow emanating from the giant radio lobes of the radio 
galaxy Centaurus~A. The resolved $\gamma$-ray image shows the lobes 
clearly separated from the central active source. In contrast to all 
other active galaxies detected so far in high-energy $\gamma$-rays, the 
lobe flux constitutes a considerable portion ($>$1/2) of the total 
source emission. The $\gamma$-ray emission from the lobes is interpreted 
as inverse Compton scattered relic radiation from the cosmic microwave 
background (CMB), with additional contribution at higher energies from 
the infrared-to-optical extragalactic background light (EBL). These 
measurements provide $\gamma$-ray constraints on the magnetic field and 
particle energy content in radio galaxy lobes, and a promising method to 
probe the cosmic relic photon fields. \end{sciabstract}

Cen~A is one of the brightest radio sources in the sky and was 
amongst the first identified with a galaxy (NGC~5128) outside of our 
Milky Way \cite{bol49}. Straddling the bright central source is a 
pair of extended radio lobes with a total angular extent of 
$\sim$10\deg\ \cite{sha58,she58}, which makes Cen~A the largest 
discrete non-thermal extragalactic radio source visible from the 
Earth. At a distance 3.7 Mpc \cite{fer07}, it is the nearest radio 
galaxy to us, and the implied physical source size is $\sim$600 kpc. 
Such double-lobed radio structures associated with otherwise 
apparently normal giant elliptical galaxies have become the defining 
feature of radio galaxies in general. The consensus explanation for 
this phenomenon is that the lobes are fueled by relativistic jets 
produced by accretion activity in a super-massive black hole 
residing at the galaxy's center.

With its unprecedented sensitivity and imaging capability (per-photon 
resolution, $\theta_{\rm 68} \simeq 0\deg.8~ E_{\rm GeV}^{-0.8}$), the 
$Fermi$ Large Area Telescope (LAT) \cite{atw09} has detected and imaged 
the radio lobes of Cen~A in high-energy $\gamma$-rays. The LAT image 
resulting from $\sim$10 months of all-sky survey data (Fig.~1) clearly 
shows the $\gamma$-ray peak coincident with the active galactic nucleus 
(AGN) detected by the {\it Compton/EGRET} instrument \cite{har99} and 
extended emission from the southern giant lobe. Because the northern 
lobe is characterized by lower surface brightness emission (in radio), 
it is not immediately apparent from a by-eye inspection of the 
$\gamma$-ray counts map. Nevertheless, from a counts profile extracted 
along the north-south axis of the source (Fig.~2) $\gamma$-ray excesses 
from both lobes are clearly visible.

Spectra for each of the lobes together with the central source 
(hereafter, the ``core'') were determined with a binned maximum 
likelihood analysis implemented in {\tt GTLIKE} \cite{cicerone} using 
events from 0.2--30 GeV in equal logarithmically spaced energy bins. 
Background emission was modeled by including the Galactic diffuse 
component, an isotropic component, and nearby $\gamma$-ray point sources 
(SOM). We fit the core as a point source at the known radio position and 
the lobe emission was modeled using a 22 GHz WMAP image (\cite{hin09}; 
Fig.~1) with the core region within a 1\deg\ radius excluded as a 
spatial template. The modeled lobe region roughly corresponds to the 
region 1 \& 2 (north), and 4 \& 5 (south) defined in \cite{har09}, where 
region 3 is the core (Fig.~2). Assuming a power-law for the $\gamma$-ray 
spectra, we find a large fraction ($>$1/2) of the total $>$100 MeV 
emission from Cen~A to originate from the lobes with the flux in each of 
the northern ((0.77 (+0.23/--0.19)$_{\rm stat.}(\pm 0.39)_{\rm syst.}) 
\times 10^{-7}$ ph cm$^{-2}$ s$^{-1}$) and southern ((1.09 
(+0.24/--0.21)$_{\rm stat.}(\pm 0.32)_{\rm syst.}) \times 10^{-7}$ ph 
cm$^{-2}$ s$^{-1}$) lobes smaller than the core flux ((1.50 
(+0.25/--0.22)$_{\rm stat.}(\pm 0.37)_{\rm syst.}) \times 10^{-7}$ ph 
cm$^{-2}$ s$^{-1}$). Uncertainties in the LAT effective area, the 
Galactic diffuse model used, and the core exclusion region were 
considered as sources of systematic error (SOM). The resultant test 
statistic (TS; \cite{mat96}) for the northern and southern giant lobes 
are 29 and 69, which corresponds to detection significances of 
$5.0\sigma$ and $8.0\sigma$, respectively. The lobe spectra are steep, 
with photon indices, $\Gamma$ = 2.52 (+0.16/--0.19)$_{\rm stat.} (\pm 
0.25)_{\rm syst.}$ (north) and 2.60 (+0.14/--0.15)$_{\rm stat.} (\pm 
0.20)_{\rm syst.}$ (south) where photons up to $\sim$2--3 GeV are 
currently detected. These values are consistent with that of the core 
($\Gamma = 2.67~(\pm 0.10)_{\rm stat.} (\pm 0.08)_{\rm syst.}$) which is 
known to have a steep $\gamma$-ray spectrum \cite{har99}. For further 
details pertaining to the analysis of the lobe emission see the SOM.

It is well-established that radio galaxy lobes are filled with 
magnetized plasma containing ultra-relativistic electrons emitting 
synchrotron radiation in the radio band (observed frequencies, $\nu \sim 
10^{7} - 10^{11}$ Hz). These electrons also up-scatter ambient photons 
to higher energies via the inverse Compton (IC) process. At the observed 
distances far from the parent galaxy ($>$100 kpc-scale), the dominant 
soft photon field surrounding the extended lobes is the pervading 
radiation from the CMB \cite{har79}. Because IC/CMB scattered emission 
in the lobes of more distant radio galaxies is generally well observed 
in the X-ray band \cite{fei95,cro05,kat05}, the IC spectrum can be 
expected to extend to even higher energies \cite{che07,har09}, as 
demonstrated by the LAT detection of the Cen~A giant lobes.

To model the observed lobe $\gamma$-rays as IC emission, detailed radio 
measurements of the lobes' synchrotron continuum spectra are necessary 
to infer the underlying electron energy distribution (EED), $n_{\rm 
e}(\gamma)$, where the electron energy is $E_{\rm e} = \gamma m_{\rm e} 
c^{2}$. In anticipation of these $Fermi$ observations, ground-based 
\cite{jun93,alv00} and WMAP satellite \cite{hin09} maps of Cen~A were 
previously analyzed \cite{har09}. Here, we separately fit the $0.4-60$ 
GHz measurements for each region defined therein for the north (1 and 2) 
and south (4 and 5) lobes (see Fig.~2) with EEDs in the form of a broken 
power-law (with normalization $k_{\rm e}$ and slopes $s_1$ and $s_2$) 
plus an exponential cutoff at high energies: $n_{\rm e}(\gamma) = k_{\rm 
e} \, \gamma^{-s_1}$ for $\gamma_{\rm min} \leq \gamma < \gamma_{\rm 
br}$ and $n_{\rm e}(\gamma) = k_{\rm e} \, \gamma_{\rm br}^{s_2-s_1} \, 
\gamma^{-s_2} \, \exp[-\gamma / \gamma_{\rm max}]$ for $\gamma \geq 
\gamma_{\rm br}$, such that the electron energy density is $U_{\rm e} = 
\int \, E_{\rm e} \, n_{\rm e}(\gamma) \, d\gamma$. To a certain extent, 
our modeling results depend on the shape of the electron spectrum at 
energies higher than those probed by the WMAP measurements 
($\nu\simgt$60 GHz; Fig.~3); we have assumed the spectrum to decline 
exponentially.

The IC spectra resulting from the fitted EED (parameters listed in 
Table~S1 of the SOM) were calculated employing precise 
synchrotron \cite{cru86} and IC \cite{blu70} kernels (including 
Klein-Nishina effects) by adjusting the magnetic field, $B$. In addition 
to the CMB photons, we included IC emission off the isotropic 
infrared-to-optical EBL radiation field \cite{hau01,geo08,har09}, 
utilizing the data compilation of \cite{rau08}. Anisotropic radiation 
from the host galaxy starlight and the well-known dust lane was also 
included, but was found to have a negligible contribution in comparison 
to the EBL (Fig.~4, see also the SOM). The resultant total IC spectra of 
the northern and southern lobes (Fig.~3) with $B = 0.89$\,$\mu$G (north) 
and $B = 0.85$\,$\mu$G (south) provide satisfactory representations of 
the observed $\gamma$-ray data. These $B$-field values imply that the 
high-energy $\gamma$-ray emission detected by the LAT is dominated by 
the scattered CMB emission, with the EBL contributing at higher energies 
($\simgt$1 GeV; Fig.~4).

Considering only contributions from ultra-relativistic electrons and 
magnetic field, the lobe plasma is found to be close to the minimum 
energy condition with the ratio of the energy densities, $U_{\rm e} 
/ U_{\rm B} \simeq 4.3$ (north) and $\simeq 1.8$ (south), where 
$U_{\rm B} = B^2 / 8 \pi$. The EED was assumed to extend down to 
$\gamma_{\rm min} = 1$; adopting larger values can reduce this ratio 
by a fractional amount for the south lobe, and up to $\sim 2 \times$ 
for the north lobe (SOM). For comparison, IC/CMB X-ray measurements 
of extended lobes of more powerful (FRII) radio sources have been 
used to infer higher $B$-fields and equipartition ratios with a 
range $U_{\rm e} / U_{\rm B} \simeq 1-10$ \cite{fei95,cro05,kat05}.

The radiating particles in the Cen~A lobes lose energy predominantly 
through the IC channel, because the ratio of the corresponding cooling 
times is equal to the energy density ratio, $U_{\rm CMB} / U_{\rm B} 
\simgt 10$. This manifests itself in the $\sim$order of magnitude 
dominance of the $\gamma$-ray component over the radio one in the 
observed SEDs (Fig.~3). However, the magnetic field constraints (thus 
the exact ratios of $U_{\rm CMB} / U_{\rm B}$) are sensitive to the 
shape of the EED at the electron energies, $E_{\rm e} > 0.1$ TeV. On one 
hand, magnetic field strengths greater than $B$$\sim$1 $\mu$G will 
underproduce the observed LAT emission for all reasonable forms of the 
EED, so the quoted ratio is formally a lower limit. Conversely, magnetic 
fields as low as $\sim$1/3 of our quoted values are strictly allowed if 
we invoke a sharper cutoff in the synchrotron spectrum at $\simgt$60 
GHz, as would be expected in some aging models for extended radio lobes 
(cf., \cite{har09}). Such models with lower magnetic fields and EEDs 
with sharper upper energy cutoffs than the exponential form adopted here 
(Fig.~3) would result in IC spectra where the EBL rather than the CMB 
component become dominant in the LAT observing band. These models 
require large departures from equipartition ($U_{\rm e} / U_{\rm B} 
\simgt 10$); even lower $B$-fields would violate the observed X-ray 
limit to the lobe flux \cite{mar81,har09}.

For a tangled magnetic field, the total non-thermal pressures in the 
lobes are, $p_{\rm rel} = (U_{\rm e} + U_{\rm B})/3$ $\simeq 5.6 \times 
10^{-14}$\,erg\,cm$^{-3}$ (north) and $\simeq 2.7 \times 
10^{-14}$\,erg\,cm$^{-3}$ (south). Such estimates can be compared to the 
ambient thermal gas pressure to enable further understanding of the 
dynamical evolution of such giant structures in general. Unfortunately, 
the parameters of the thermal gas at the appropriate distances from the 
nucleus of Cen A are not well known. Upper limits of the soft X-ray 
emission of the lobes \cite{har09}, as well as Faraday rotation studies 
\cite{fea09} indicate that the thermal gas number density is $n_{\rm 
gas} < 10^{-4}$\,cm$^{-3}$ within the giant lobes. Hence, the upper 
limit for the thermal pressure, $p_{\rm gas} = n k T < 10^{-13} \, 
(n_{\rm gas}/10^{-4}\,{\rm cm^{-3}})\, (T_{\rm gas}/10^7\,{\rm 
K})$\,erg\,cm$^{-3}$, is comparable to the evaluated non-thermal 
pressures.

Our modeling results allow us to estimate the total energy in both giant 
lobes, $E_{\rm tot} \simeq 1.5 \times 10^{58}$\,erg. This, divided by 
the lifetime of the lobes derived from spectral aging \cite{har09}, 
$\tau \simeq 3 \times 10^7$\,yrs, gives the required kinetic power of 
the jets inflating the giant lobes, $L_{\rm j} \simeq E_{\rm tot} / 
2\,\tau \simeq 7.7 \times 10^{42}$\,erg\,s$^{-1}$, which is close to the 
estimates of the total power of the kpc-scale outflow in the current 
epoch of jet activity \cite{cro09}. For a black hole mass in Cen A, 
$M_{\rm BH} \simeq 10^8\,M_{\odot}$ \cite{mar06}, this implies a jet 
power which is only a small fraction of the Eddington luminosity 
($L_{\rm j} \simeq 6.1 \times 10^{-4} L_{\rm Edd}$), and a relatively 
small jet production efficiency, $E_{\rm tot} / M_{\rm BH} c^2 \simeq 8 
\times 10^{-5}$. Because the $p\,dV$ work done by the expanding lobes on 
the ambient medium is not taken into account, and the relativistic 
proton content is unconstrained in our analysis, the obtained values for 
$E_{\rm tot}$ and $L_{\rm j}$ are strict lower limits and could 
plausibly be an order of magnitude larger (cf., \cite{der09}).

The observed LAT emission implies the presence of 0.1--1 TeV electrons 
in the $>100$'s kpc-scale lobes. Because their radiative lifetimes ($<$ 
1--10 Myr) approach plausible electron transport timescales across the 
lobes, the particles have either been accelerated in situ or efficiently 
transported from regions closer to the nucleus. Such high-energy 
electrons in the lobes are in fact required to IC scatter photons into 
the LAT band and it is presently unclear how common this is in other 
radio galaxies.


\begin{figure}[htbp]
  \begin{center}
    \includegraphics[width=7.5cm]{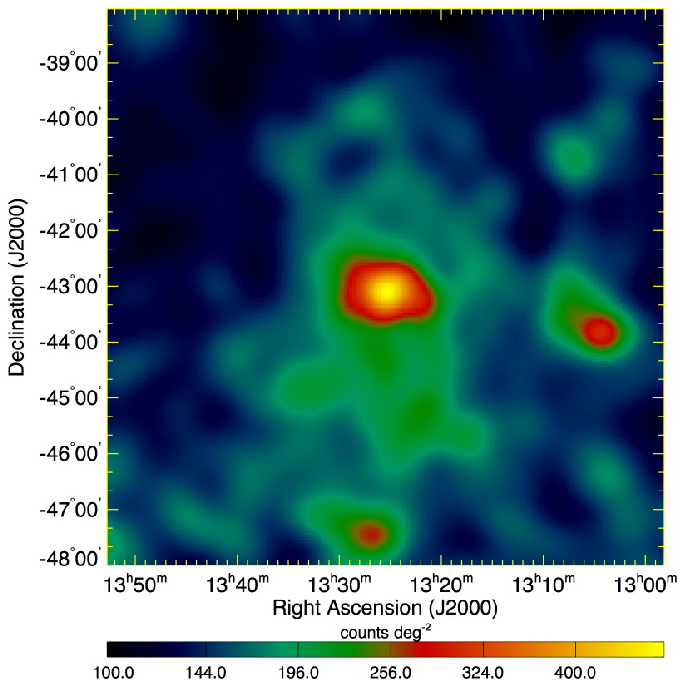}
    \includegraphics[width=7.5cm]{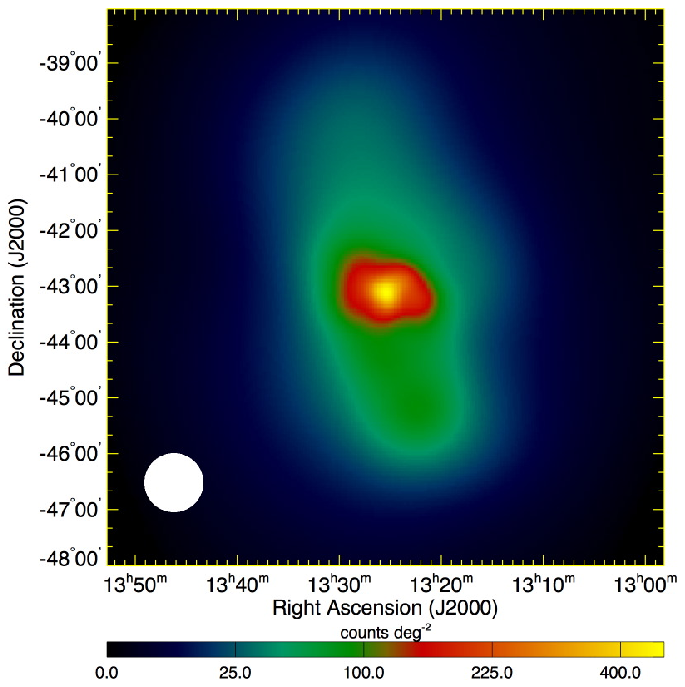}
    \includegraphics[width=7.5cm]{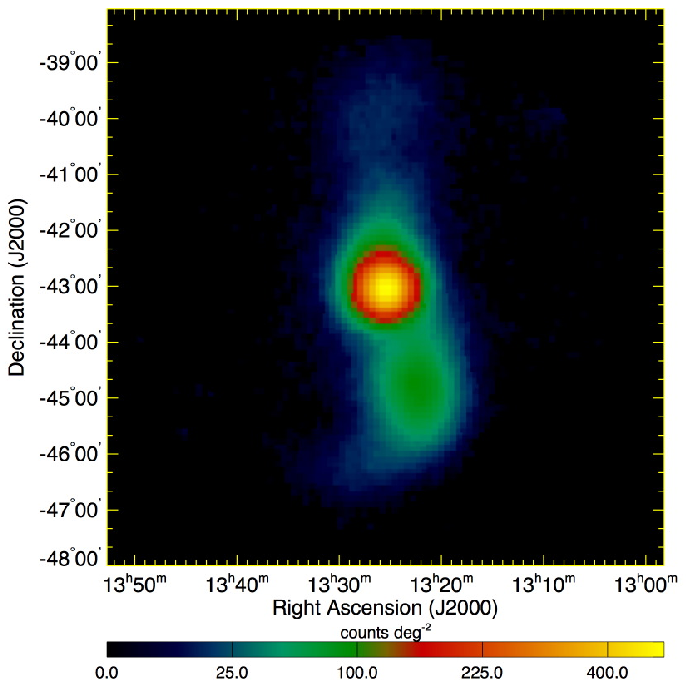}
  \end{center} 
{\bf Fig.~1.} $Fermi$-LAT $\gamma$-ray ($>$200 MeV) counts maps
centered on Cen~A displayed with square-root scaling {\bf (A, B)}. In both panels, a
model of the Galactic and isotropic emission components were subtracted
(in contrast to the observed counts profile presented in Fig.~2). The
images are shown before {\bf (A)} and and after {\bf (B)} additional
subtraction of field point sources (SOM), and are shown adaptively
smoothed with a minimum signal-to-noise ratio of 10.  In
panel {\bf (B)}, the white circle with a diameter of 1\deg\ is approximately
the scale of the LAT PSF width.
For comparison, the 22 GHz radio map from the 5-year WMAP
dataset \cite{hin09} with a resolution 0\deg.83 is shown {\bf (C)}.
\end{figure}

\begin{figure}[htbp]
  \begin{center}
    \includegraphics[width=16cm]{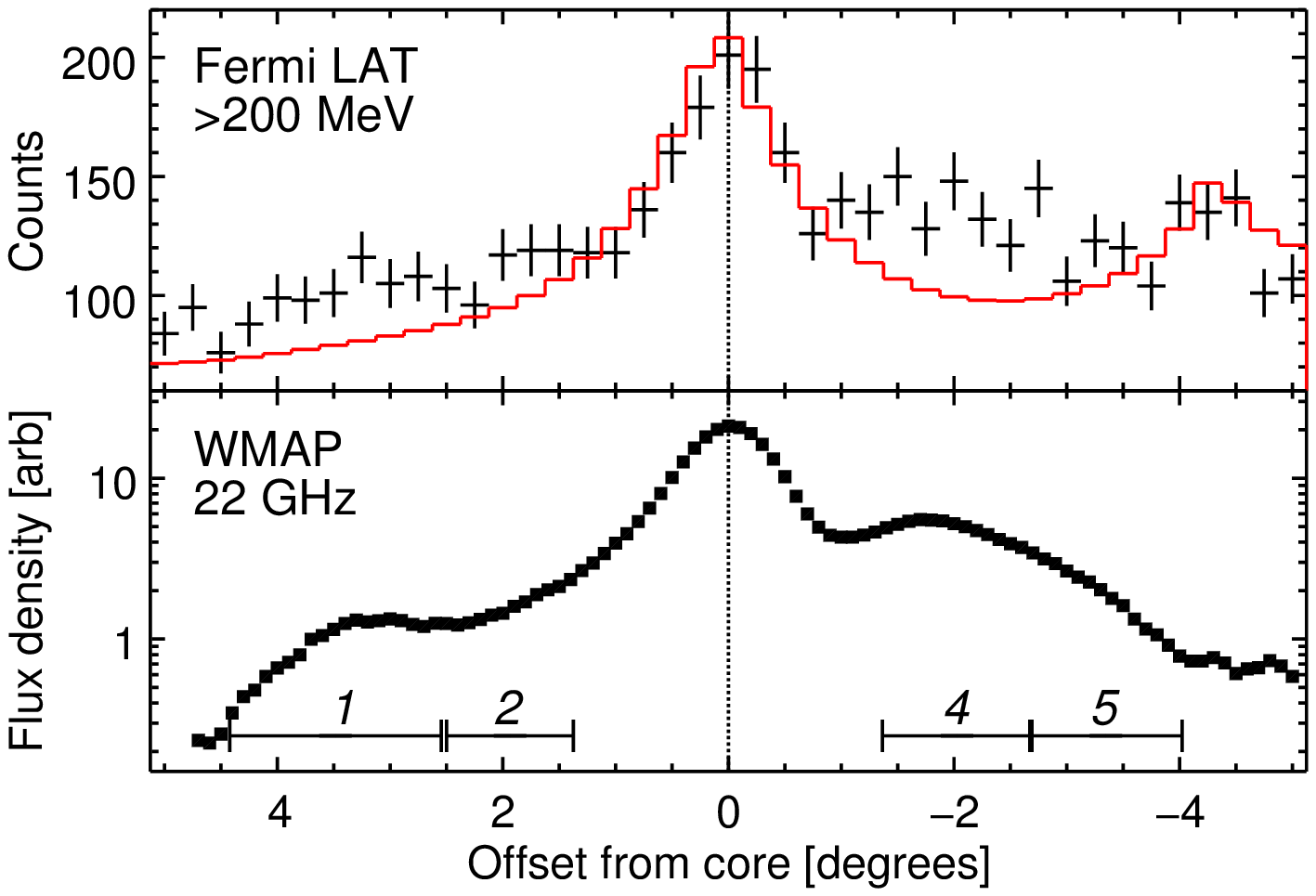}
  \end{center}
{\bf Fig.~2.} Observed intensity profiles of Cen~A along the north-south 
axis in $\gamma$-rays [top panel] and radio [bottom panel]. In the 
bottom panel, the lobe regions 1 \& 2 (northern lobe), and 4 \& 5 
(southern lobe) are indicated following \cite{har09} where region 3 (not 
indicated) is the core. The red curve overlaid onto the LAT data 
indicates the emission model for of all fitted points sources plus the 
isotropic and Galactic diffuse (brighter to the south) emission. The 
point sources include the Cen~A core (offset=0\deg) and a LAT source at 
an offset=$-$4.5\deg\ (see SOM) which is clearly outside (1\deg\ from 
the southern edge) of the southern lobe. The excess counts are 
coincident with the northern and southern giant lobes. \end{figure}

\begin{figure}[htbp]
  \begin{center}
    \includegraphics[width=7.75cm]{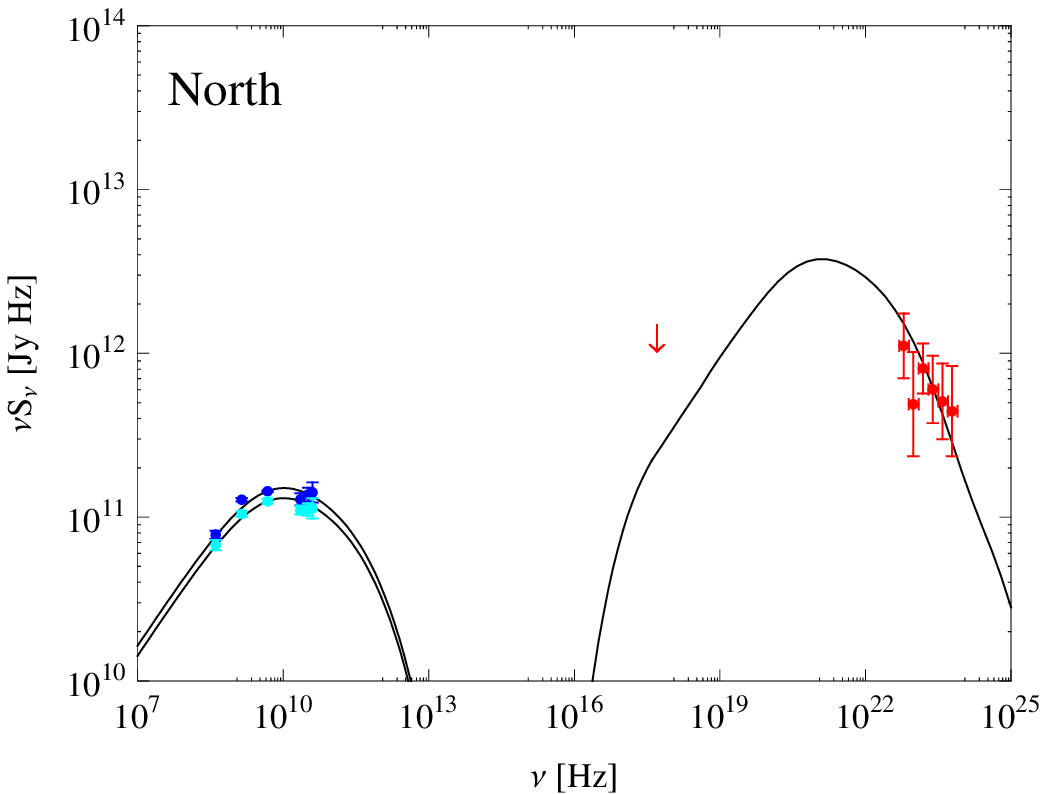}
    \includegraphics[width=7.75cm]{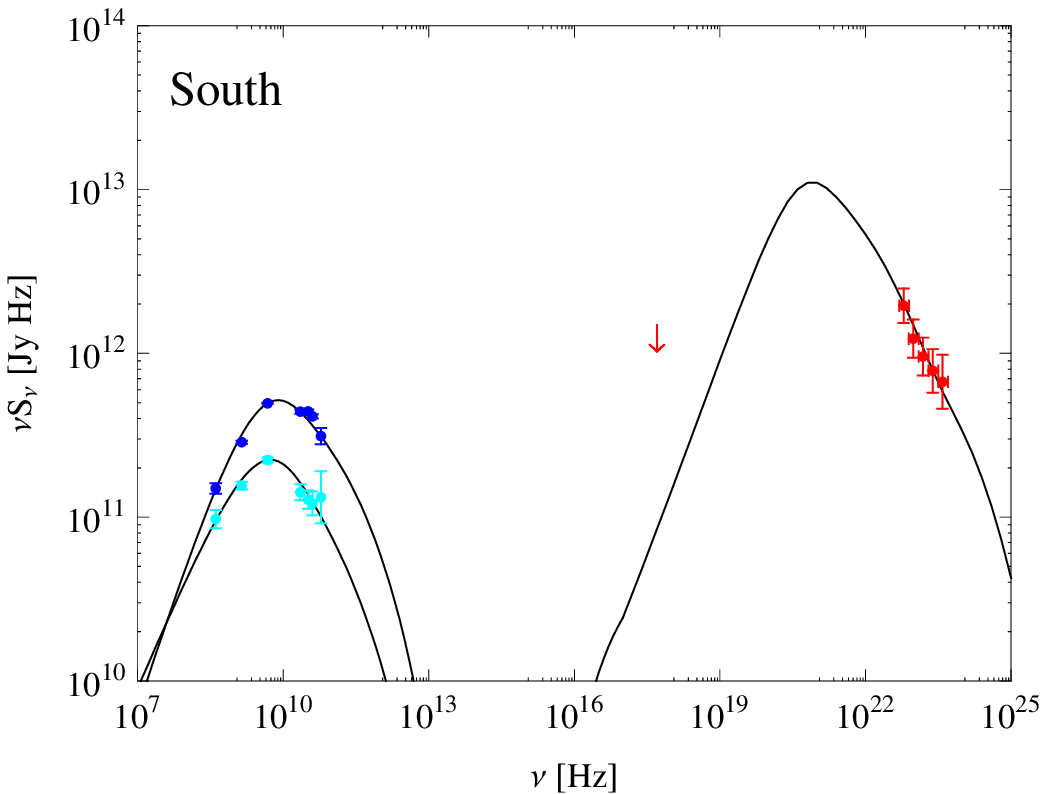}
  \end{center}
{\bf Fig.~3.} Broad-band spectral energy distributions (SEDs) of the
northern {\bf (A)} and southern {\bf (B)} giant lobes of Cen~A.
The radio measurements (up 
to 60 GHz) of each lobe are separated into 2 regions with blue data 
points indicating regions (2 and 4; cf., Fig.~2) closer to the nucleus 
and the farther regions (1 and 5) in light blue. Synchrotron continuum 
models for each region are overlaid (see text). The component at higher 
energies is the total IC emission of each lobe modeled to match the LAT 
measurements (red points with error bars). The X-ray limit for the lobe 
emission derived from $SAS$-3 observations \cite{mar81} is indicated 
with a red arrow; see \cite{har09}. The break and maximum frequencies in 
the synchrotron spectra are $\nu_{\rm br} = 4.8$ GHz and $\nu_{\rm max} 
= 400$ GHz, respectively. \end{figure}

\begin{figure}[htbp]
  \begin{center}
    \includegraphics[width=7.75cm]{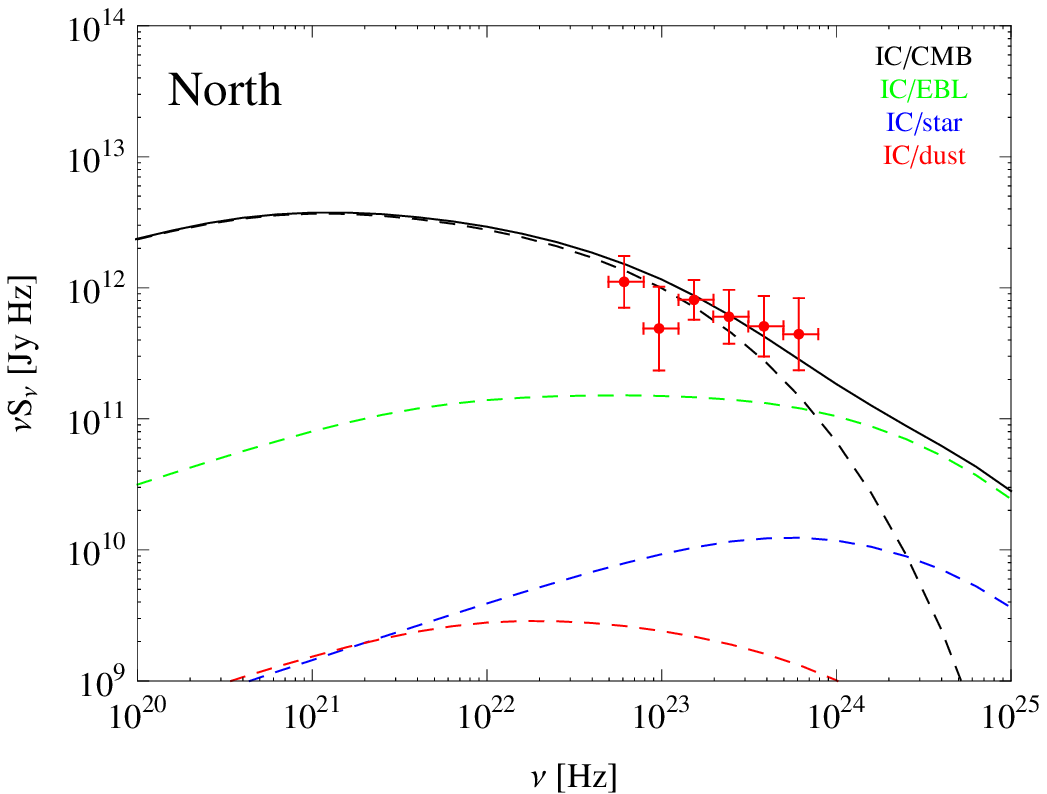}
    \includegraphics[width=7.75cm]{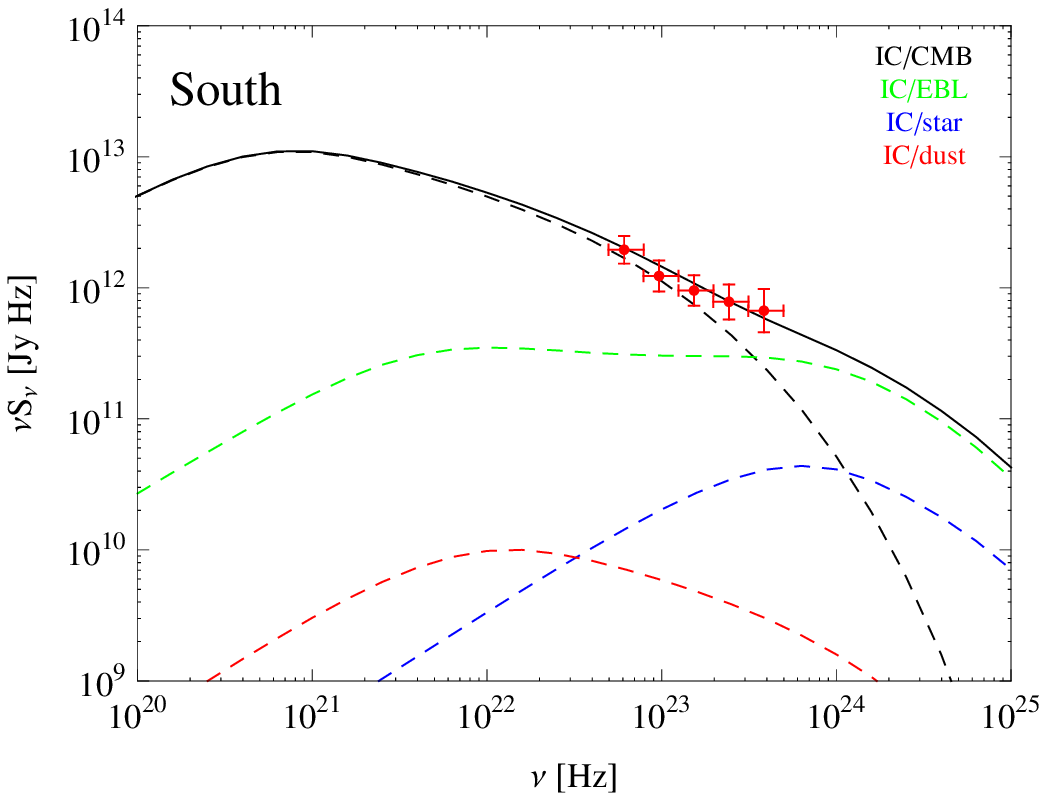}
  \end{center}
{\bf Fig.~4.} Detail of the IC portion of the northern {\bf (A)} and southern {\bf (B)}
giant lobes' SEDs (cf., Fig.~3). The separate contributions from the 
different photon seed sources are indicated with dashed lines while the 
total emission is the solid black line. \end{figure}

\clearpage

\noindent {\bf The Fermi LAT Collaboration}\\
\noindent
A.~A.~Abdo$^{1,2}$, 
M.~Ackermann$^{3}$, 
M.~Ajello$^{3}$, 
W.~B.~Atwood$^{4}$, 
L.~Baldini$^{5}$, 
J.~Ballet$^{6}$, 
G.~Barbiellini$^{7,8}$, 
D.~Bastieri$^{9,10}$, 
B.~M.~Baughman$^{11}$, 
K.~Bechtol$^{3}$
% ...
\medskip
\begin{enumerate}
\item[1.] Space Science Division, Naval Research Laboratory, Washington, DC 20375, USA
\item[2.] National Research Council Research Associate, National Academy of Sciences, Washington, DC 20001, USA
\item[3.] W. W. Hansen Experimental Physics Laboratory, Kavli Institute for Particle Astrophysics and Cosmology, Department of Physics and SLAC National Accelerator Laboratory, Stanford University, Stanford, CA 94305, USA
\item[4.] Santa Cruz Institute for Particle Physics, Department of Physics and Department of Astronomy and Astrophysics, University of California at Santa Cruz, Santa Cruz, CA 95064, USA
\item[5.] Istituto Nazionale di Fisica Nucleare, Sezione di Pisa, I-56127 Pisa, Italy
\item[6.] Laboratoire AIM, CEA-IRFU/CNRS/Universit\'e Paris Diderot, Service d'Astrophysique, CEA Saclay, 91191 Gif sur Yvette, France
\item[7.] Istituto Nazionale di Fisica Nucleare, Sezione di Trieste, I-34127 Trieste, Italy
\item[8.] Dipartimento di Fisica, Universit\`a di Trieste, I-34127 Trieste, Italy
\item[9.] Istituto Nazionale di Fisica Nucleare, Sezione di Padova, I-35131 Padova, Italy
\item[10.] Dipartimento di Fisica ``G. Galilei", Universit\`a di Padova, I-35131 Padova, Italy
\item[11.] Department of Physics, Center for Cosmology and Astro-Particle Physics, The Ohio State University, Columbus, OH 43210, USA
\item[12.] Istituto Nazionale di Fisica Nucleare, Sezione di Perugia, I-06123 Perugia, Italy
\item[13.] Dipartimento di Fisica, Universit\`a degli Studi di Perugia, I-06123 Perugia, Italy
\item[14.] Dipartimento di Fisica ``M. Merlin" dell'Universit\`a e del Politecnico di Bari, I-70126 Bari, Italy
\item[15.] Istituto Nazionale di Fisica Nucleare, Sezione di Bari, 70126 Bari, Italy
\item[16.] Laboratoire Leprince-Ringuet, \'Ecole polytechnique, CNRS/IN2P3, Palaiseau, France
\item[17.] Department of Physics, University of Washington, Seattle, WA 98195-1560, USA
\item[18.] Institut de Ciencies de l'Espai (IEEC-CSIC), Campus UAB, 08193 Barcelona, Spain
\item[19.] INAF-Istituto di Astrofisica Spaziale e Fisica Cosmica, I-20133 Milano, Italy
\item[20.] Agenzia Spaziale Italiana (ASI) Science Data Center, I-00044 Frascati (Roma), Italy
\item[21.] NASA Goddard Space Flight Center, Greenbelt, MD 20771, USA
\item[22.] Center for Research and Exploration in Space Science and Technology (CRESST) and NASA Goddard Space Flight Center, Greenbelt, MD 20771, USA
\item[23.] Department of Physics and Center for Space Sciences and Technology, University of Maryland Baltimore County, Baltimore, MD 21250, USA
\item[24.] George Mason University, Fairfax, VA 22030, USA
\item[25.] Laboratoire de Physique Th\'eorique et Astroparticules, Universit\'e Montpellier 2, CNRS/IN2P3, Montpellier, France
\item[26.] Department of Physics and Astronomy, Sonoma State University, Rohnert Park, CA 94928-3609, USA
\item[27.] Department of Physics, Stockholm University, AlbaNova, SE-106 91 Stockholm, Sweden
\item[28.] The Oskar Klein Centre for Cosmoparticle Physics, AlbaNova, SE-106 91 Stockholm, Sweden
\item[29.] Royal Swedish Academy of Sciences Research Fellow, funded by a grant from the K. A. Wallenberg Foundation
\item[30.] Dipartimento di Fisica, Universit\`a di Udine and Istituto Nazionale di Fisica Nucleare, Sezione di Trieste, Gruppo Collegato di Udine, I-33100 Udine, Italy
\item[31.] Universit\'e de Bordeaux, Centre d'\'Etudes Nucl\'eaires Bordeaux Gradignan, UMR 5797, Gradignan, 33175, France
\item[32.] CNRS/IN2P3, Centre d'\'Etudes Nucl\'eaires Bordeaux Gradignan, UMR 5797, Gradignan, 33175, France
\item[33.] Department of Physical Sciences, Hiroshima University, Higashi-Hiroshima, Hiroshima 739-8526, Japan
\item[34.] Department of Astronomy and Astrophysics, Pennsylvania State University, University Park, PA 16802, USA
\item[35.] Department of Physics and Department of Astronomy, University of Maryland, College Park, MD 20742, USA
\item[36.] INAF Istituto di Radioastronomia, 40129 Bologna, Italy
\item[37.] Max-Planck-Institut f\"ur Radioastronomie, Auf dem H\"ugel 69, 53121 Bonn, Germany
\item[38.] Center for Space Plasma and Aeronomic Research (CSPAR), University of Alabama in Huntsville, Huntsville, AL 35899, USA
\item[39.] Department of Physics, Royal Institute of Technology (KTH), AlbaNova, SE-106 91 Stockholm, Sweden
\item[40.] Waseda University, 1-104 Totsukamachi, Shinjuku-ku, Tokyo, 169-8050, Japan
\item[41.] Department of Physics, Tokyo Institute of Technology, Meguro City, Tokyo 152-8551, Japan
\item[42.] Cosmic Radiation Laboratory, Institute of Physical and Chemical Research (RIKEN), Wako, Saitama 351-0198, Japan
\item[43.] Centre d'\'Etude Spatiale des Rayonnements, CNRS/UPS, BP 44346, F-30128 Toulouse Cedex 4, France
\item[44.] Istituto Nazionale di Fisica Nucleare, Sezione di Roma ``Tor Vergata", I-00133 Roma, Italy
\item[45.] Department of Physics and Astronomy, University of Denver, Denver, CO 80208, USA
\item[46.] Max-Planck Institut f\"ur extraterrestrische Physik, 85748 Garching, Germany
\item[47.] Institut f\"ur Astro- und Teilchenphysik and Institut f\"ur Theoretische Physik, Leopold-Franzens-Universit\"at Innsbruck, A-6020 Innsbruck, Austria
\item[48.] Space Sciences Division, NASA Ames Research Center, Moffett Field, CA 94035-1000, USA
\item[49.] NYCB Real-Time Computing Inc., Lattingtown, NY 11560-1025, USA
\item[50.] Astronomical Observatory, Jagiellonian University, 30-244 Krak\'ow, Poland
\item[51.] Department of Chemistry and Physics, Purdue University Calumet, Hammond, IN 46323-2094, USA
\item[52.] Institute of Space and Astronautical Science, JAXA, 3-1-1 Yoshinodai, Sagamihara, Kanagawa 229-8510, Japan
\item[53.] Instituci\'o Catalana de Recerca i Estudis Avan\c{c}ats (ICREA), Barcelona, Spain
\item[54.] Consorzio Interuniversitario per la Fisica Spaziale (CIFS), I-10133 Torino, Italy
\item[55.] Dipartimento di Fisica, Universit\`a di Roma ``Tor Vergata", I-00133 Roma, Italy
\item[56.] School of Pure and Applied Natural Sciences, University of Kalmar, SE-391 82 Kalmar, Sweden
\item[57.] Centre for Astrophysics Research, University of Hertfordshire, College Lane, Hatfield AL10 9AB, UK
\item[$\dagger$] To whom correspondence should be addressed.\\ 
E-mail: Teddy.Cheung.ctr@nrl.navy.mil (C.C.C.); fukazawa@hep01.hepl.hiroshima-u.ac.jp (Y.F.); 
jurgen.knodlseder@cesr.fr (J.K.); stawarz@slac.stanford.edu (\L.S.)

\end{enumerate}

\newpage
\setcounter{page}{1}


\section*{Supporting Online Material (SOM)}

\section*{Materials and Methods: LAT Data Selection and Tests of Spatial 
Modeling}

The characteristics and performance of the LAT aboard {\em Fermi} are 
described in detail by \cite{atw09s}. The data used in this work amount 
to 300 days of continuous sky survey observations over the period August 
4th 2008 -- May 31st 2009 (corresponding to mission elapsed times (MET) 
239557420 - 265507200) during which an effective exposure of $\sim 2.3 
\times 10^{10}$~cm$^2$~s (at 1 GeV) is obtained for Cen~A. Events 
satisfying the standard low-background event selection (``Pass 6 Diffuse'' 
events) \cite{atw09s}, coming from zenith angles $<105^{\circ}$ and 
satisfying the rocking angle cut of $39^{\circ}$ are used 
\cite{abdo09s}. We further restrict the analysis to photon energies 
above 200~MeV; below this energy the effective area in the Diffuse class 
is relatively small and strongly dependent on energy.

For the analysis we select all events within a rectangular 
region-of-interest (ROI) of size $14^{\circ} \times 14^{\circ}$ centered 
on $(\alpha_{\rm J2000}, \delta_{\rm J2000})=(13^{\rm h}25^{\rm 
m}26^{\rm s}, -43^\circ01'12")$ and aligned in equatorial coordinates. A 
counts map of the ROI is shown in Fig.~S1. All analysis is performed 
using the LAT Science Tools package, which is available from the Fermi 
Science Support Center \cite{cicerones}, using P6\_V3 post-launch 
instrument response functions (IRFs). These take into account pile-up 
and accidental coincidence effects in the detector subsystems that were 
not considered in the definition of the pre-launch IRFs. At the Galactic 
latitude of Cen~A ($b \approx +19^{\circ}$), the $\gamma$-ray background 
is a combination of extragalactic and Galactic diffuse emissions and 
some residual instrumental background. We model the Galactic background 
component using the LAT standard diffuse background model {\tt 
gll\_iem\_v02}\footnote{The model can be downloaded from 
http://fermi.gsfc.nasa.gov/ssc/data/access/lat/BackgroundModels.html.} 
for which we keep the overall normalization as a free parameter. The 
extragalactic and residual instrumental backgrounds are combined into a 
single component which has been taken as being isotropic and with a 
spectral distribution determined from all-sky fitting (file {\tt 
isotropic\_iem\_v02.txt}). The normalization of this component was left 
as a free parameter.

The first {\it Fermi}-LAT source catalog \cite{1fgl} contains 14 point 
sources within the ROI. One of these sources (1FGL~J1325.6$-$4300) 
corresponds to the core of Cen A while two sources (1FGL~J1322.0$-$4515, 
1FGL~J1333.4$-$4036) are likely local maxima of the Cen A lobes. This 
leaves 11 point sources within the ROI that are not associated to Cen A. 
To correctly account for these 11 field sources in our analysis, we 
include point sources at the positions quoted in Table~S2
in our background model for which we left the fluxes and spectral 
power-law indices as free parameters. The locations of the point sources 
are indicated by boxes in Fig.~S1.

We model the core emission from Cen A with a point source located at the 
known radio position of $(\alpha_{\rm J2000}, \delta_{\rm 
J2000})=(13^{\rm h}25^{\rm m}27^{\rm s}, -43^\circ01'09")$ for which we 
keep the flux and power-law spectral index as free parameters. The giant 
radio lobes of Cen A are modeled using a spatial template that is based 
on the WMAP 22 GHz image \cite{hin09s} of Cen A (see main text and 
Fig.~S1). To exclude the core emission from this template we set all 
pixels within a radius of $1^\circ$ around the core of Cen A to zero. We 
further split the template along an east-west axis running through the 
core of Cen A to obtained separate models for the northern and southern 
radio lobes.  We model the spectrum of both lobes using power laws for 
which we keep the fluxes and spectral indices as free parameters.

We fit the core and radio lobes in addition to the background model to 
the data using a binned maximum likelihood optimization. The test 
statistic (TS) \cite{mat96s} of the core amounts to 219, corresponding 
to a detection significance of $14.6\sigma$. For the northern and 
southern lobes we obtain TS values of 29 and 69, which corresponds to 
detection significances of $5.0\sigma$ and $8.0\sigma$, respectively. As 
a test, fake WMAP image templates were created by rotating the map by 
$90^\circ$, $180^\circ$, and $270^\circ$, bringing the lobes out of the 
alignment with respect to the radio emission. As expected, when fitting 
the data with these fake maps instead of the original WMAP template the 
TS is reduced by a large factor ($\sim4$), indicating that the 
$\gamma$-ray emission indeed matches the radio morphology.

As Cen~A is relatively close to the Galactic plane, an accurate modeling 
of the diffuse Galactic emission is important. We tested the stability 
of our results by replacing the standard model of diffuse Galactic 
emission by the model {\tt gll\_iem\_v01}, i.e., from the GALPROP code 
\cite{abdo09s} and this is considered as a source of systematic 
uncertainty. The effect of different core exclusion radii was also 
tested, considering values of 1.5\deg\ and 1.25\deg\ in addition to the 
1\deg\ results quoted. The lower end of this range is defined by 
excluding enough of the core to sufficiently model its emission, and 
the upper end is the radius in which the lobe emission begins to be
excluded. Lastly, systematic uncertainties were calculated considering 
the effect of the uncertainty in the IRFs (i.e., the LAT effective area, 
$A_{\rm eff}$). These are summarized in Table~S3.

To investigate whether the $\gamma$-ray emission seen towards the giant 
radio lobes of Cen A could also be attributed to background blazars in 
the area, we have shown in the right panel of Fig.~S1 the distribution 
of blazar candidates in the CRATES catalog of flat-spectrum radio 
sources \cite{hea07s} within the ROI. We found a total of 11 
$\gamma$-ray blazar candidates that spatially overlap with the radio 
lobes of Cen A. From Fig.~S1 we can already notice that none of them 
coincides with a local maximum of $\gamma$-ray emission in the counts 
map. Fitting point sources at the positions of the 11 CRATES sources 
with flux and spectral power law index set free in addition to the core 
and lobe model of Cen A does not result in a significant detection for 
any of them, and does not lead to a significant reduction of the flux 
from the WMAP template that models the $\gamma$-ray emission from the 
lobes. We thus conclude that the emission seen towards the Cen A radio 
lobes cannot be attributed to known blazars in the field. Extended 
emission from the giant radio lobes of Cen A seems thus as the most 
plausible explanation of the observations.

\section*{Theoretical Modeling}

Following \cite{sta06s}, we used a template elliptical galaxy spectrum 
\cite{sil98s} normalized to the $V$-band apparent magnitude, $m_V = 7.0$ 
\cite{isr98s} to derive a luminosity of the host of $L_V \simeq 7.8 
\times 10^{43}$\,erg\,s$^{-1}$ in this band. The far-infrared (FIR) 
emission of the dust lane is modeled as a modified blackbody, $\nu I 
_{\nu}(T_{\rm d}) \propto \nu^4 \, (1 - \exp[ - (\nu / \nu_0)^{-\beta}]) 
/ (\exp[h\nu / kT_d]-1 )$, where $\beta = 1.5$ is the dust emissivity 
power-law, $T_{\rm d} \simeq 60$\,K is the assumed dust temperature 
(corresponding to the peak frequencies, $\nu_{\rm d} = 3 k T_{\rm d} / h 
\simeq 3.75 \times 10^{12}$\,Hz), and $\nu_0 = 3 \times 10^{12}$\,Hz is 
the frequency below which the thermal dust emission becomes optically 
thin. This distribution is normalized to the total $100$\,$\mu$m Cen A 
flux of $400$\,Jy \cite{gol88s}, amounting to a luminosity of $L_{100} 
\simeq 2 \times 10^{43}$\,erg\,s$^{-1}$ at this wavelength. The spectra 
generated in this way are consistent with the monochromatic flux density 
measurements compiled by \cite{cro09s}.

In Fig.~S2, the relevant photon fields as seen by the lobes are plotted. 
These include the CMB (black) and the solid green line denotes the EBL 
model from \cite{rau08s} adopted in our calculations (see main text); 
green dashed lines are the other EBL models considered (see Fig.~S3). 
The volume averaged energy density of the starlight (blue) and dust 
(red) emission as seen at the locations of regions 1 and 2 (lower and 
upper solid, respectively) and 4 and 5 (upper and lower dashed, 
respectively) are indicated. For all EBL choices, the EBL energy density 
dominates over that of the host galaxy starlight and dust emission by 
$\simgt$3--100$\times$. Here, the lobes are assumed to lie in the plane 
of the sky. This assumption maximizes the possible contribution of the 
galactic emission as smaller angles to the line of sight result in 
larger de-projected distances for the radio lobes and would make the 
galactic photon fields even less relevant. In terms of our modeling 
results, smaller angles would also imply larger source volumes which 
would affect only the jet powers evaluated (making them larger), so the 
quoted estimate should further be considered as a lower limit.

As mentioned in the main text, we parameterize the electron energy 
distribution (EED) of each lobe region with a broken power-law in the 
form: $n_{\rm e}(\gamma) = k_{\rm e} \, \gamma^{-s_1}$ for $\gamma_{\rm 
min} \leq \gamma < \gamma_{\rm br}$ and $n_{\rm e}(\gamma) = k_{\rm e} 
\, \gamma_{\rm br}^{s_2-s_1} \, \gamma^{-s_2} \, \exp[-\gamma / 
\gamma_{\rm max}]$ for $\gamma \geq \gamma_{\rm br}$. We initially 
considered a wide range of magnetic field strengths, $B = 0.1-10 \mu$G, 
and adjusted the parameters of the EED ($k_{\rm e}$, $s_{\rm 1}$, $s_{\rm 
2}$, $\gamma_{\rm br}$, and $\gamma_{\rm max}$) to match the radio 
measurements. The IC fluxes were evaluated separately for each defined 
region (1, 2, 4, and 5) and the summed computed emission from each lobe 
was compared to the observed LAT spectra. The models found to best 
represent the radio and $\gamma$-ray measurements are reported (Table~S1). 
Using different EBL models/compilations (Fig.~S3), we found the 
resultant IC SEDs are insensitive to the particular EBL models within 
the measurement uncertainties.

The low-energy ultrarelativistic electron energy distributions in 
the south lobe are relatively flat, with indices $s_{\rm 1} < 2$. We 
note in this context that since the synchrotron continua for this 
lobe are significantly curved, the precise synchrotron kernel has to 
be used in evaluating the synchrotron emission for a given form of 
the electron spectrum. That is, no standard approximation relating 
the synchrotron spectral index, $\alpha$ (defined as $S_{\nu} 
\propto \nu^{-\alpha}$), with the electron energy index, $s=2 \alpha 
+1$, can be made. Such an approximation holds when the electron 
energy distribution is of the form of a single (and steep) 
power-law, but not if the electron spectrum has a sharp break, which 
-- in the considered case -- affects the synchrotron spectra in the GHz range for the Cen A south 
lobe. If we increase $s_{\rm 1}$ for the southern lobe up to 2, we 
would not be able to reproduce the radio data well. We also note 
that with $s_{\rm 1} < 2$, the derived total electron energy density 
($U_{\rm e}$) of the southern lobe is relatively insensitive to our 
assumption of the minimum electron energy ($\gamma_{\rm min} = 1$) 
with a decrease by only a few percent if we increase $\gamma_{\rm 
min}$ to 100 for example. For the north lobe however, increasing 
$\gamma_{\rm min}$ to 100 would result in a non-negligible ($\simeq 
2 \times$) decrease in $U_{\rm e}$ because this lobe is characterized by a 
steeper low-energy electron spectrum.

In addition to the galactic emission, one can expect that the 
non-thermal emission from the active nucleus also illuminates the giant 
lobes, thus providing another possible seed photon source. For Cen~A, 
the energy density of the nuclear emission at a distance $r \sim 100$ 
kpc from the nucleus (corresponding to regions 2 and 4) can be estimated 
roughly as, $U_{\rm nuc} \simlt 10^{-14}$ erg cm$^{-3}$ (see 
\cite{sta03s}). This is comparable at best to the EBL level at 
near-infrared/optical wavelengths but as the precise level is dependent 
on the unknown relativistic beaming parameters of the inner jet and 
uncertainty in the jet duty cycle, this is omitted in our estimates for 
simplicity.


\begin{figure}[htbp]
  \begin{center}
	\includegraphics[width=7.75cm]{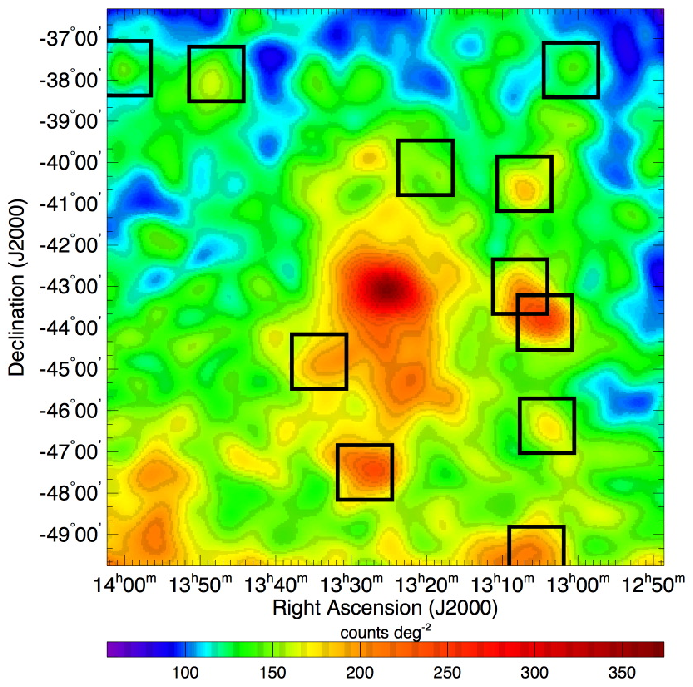}
	\includegraphics[width=7.75cm]{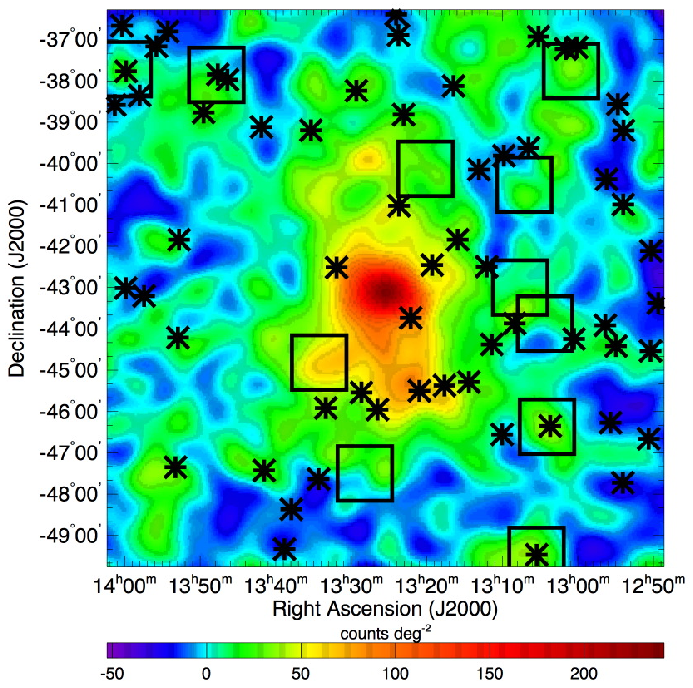}
  \end{center}
{\bf Fig.~S1.} Gaussian kernel ($\sigma=0.3^\circ$) smoothed counts maps 
of the region-of-interest (ROI) around Cen~A in a true local projection 
before (left) and after subtraction of the background model (right) for 
the energy range 200 MeV -- 30 GeV and for a pixel size of $0.1^\circ 
\times 0.1^\circ$. The boxes show the locations of the 11 LAT point 
sources that have been included in the background model. The stars in 
the right panel show the locations of the CRATES radio sources in the 
ROI. \end{figure}

\begin{figure}[htbp]
  \begin{center}
    \includegraphics[width=9cm]{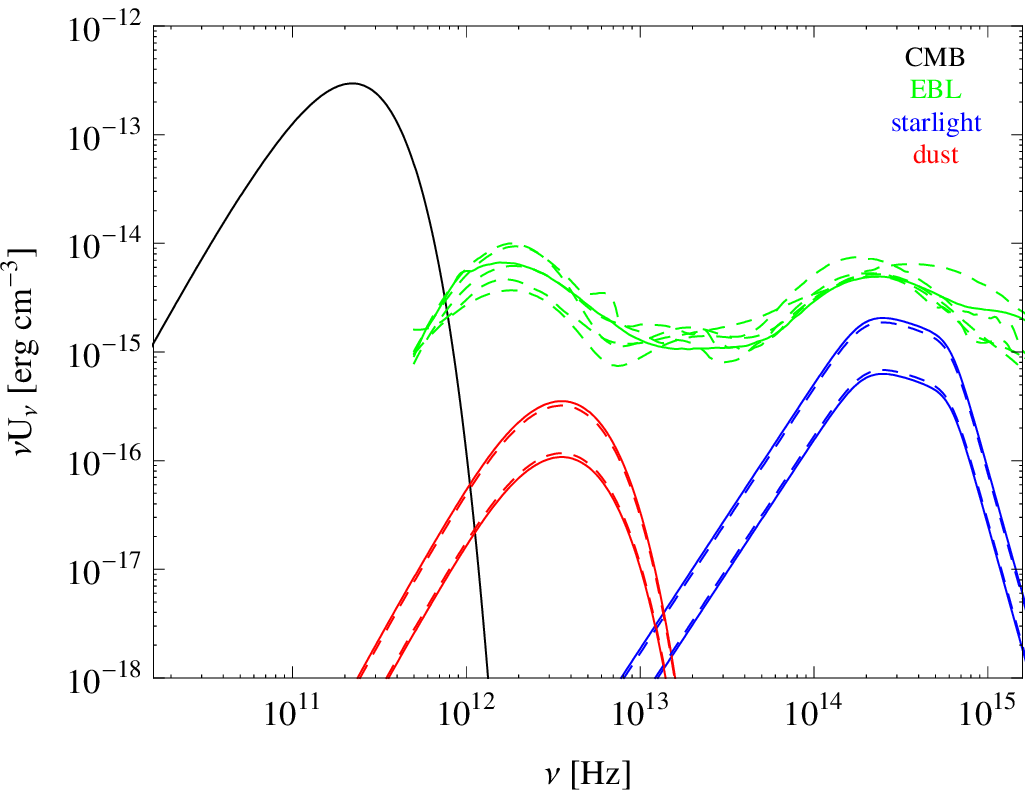}
  \end{center}
{\bf Fig.~S2.} Energy densities of different photon fields at the 
locations of the northern and southern giant lobes. For the EBL, the 
solid line indicates the compilation of \cite{rau08s} utilized for the 
IC spectra in Fig.~3 of the main paper while the dashed lines are other 
EBL models considered (see Fig.~S3). For the dust and starlight 
components, the solid lines are indicative of the northern regions (1 = 
lower, 2 = upper) and the dashed lines represent the southern regions (4 
= upper, 5 = lower). \end{figure}

\begin{figure}[htbp]
  \begin{center}
    \includegraphics[width=7.75cm]{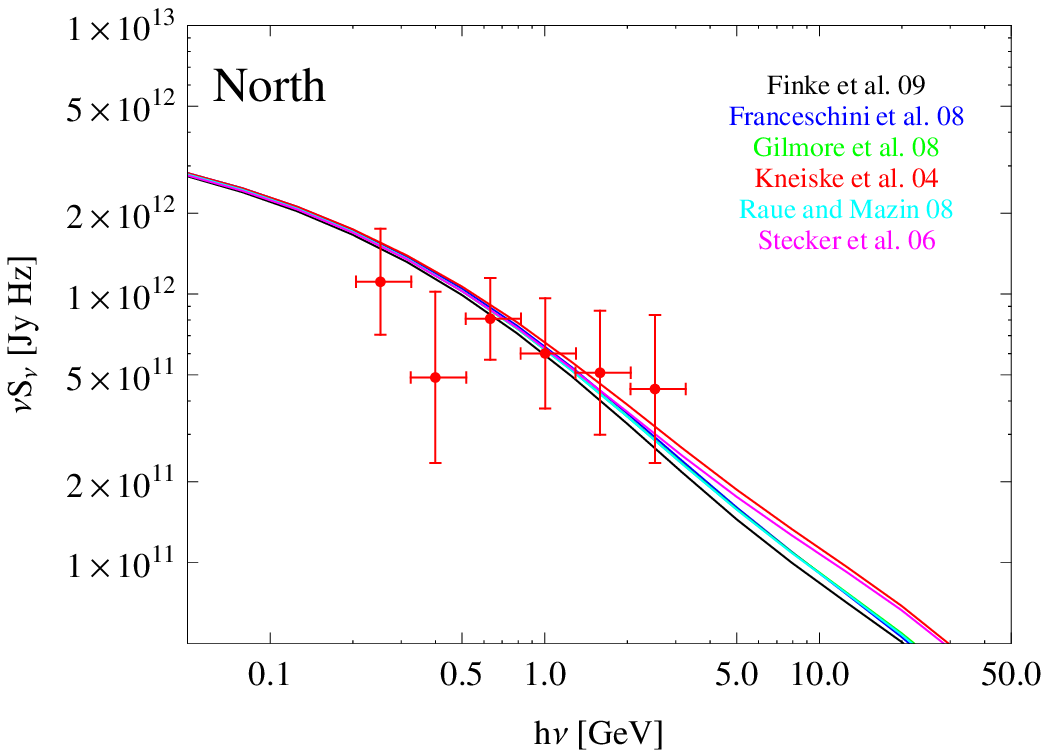}
    \includegraphics[width=7.75cm]{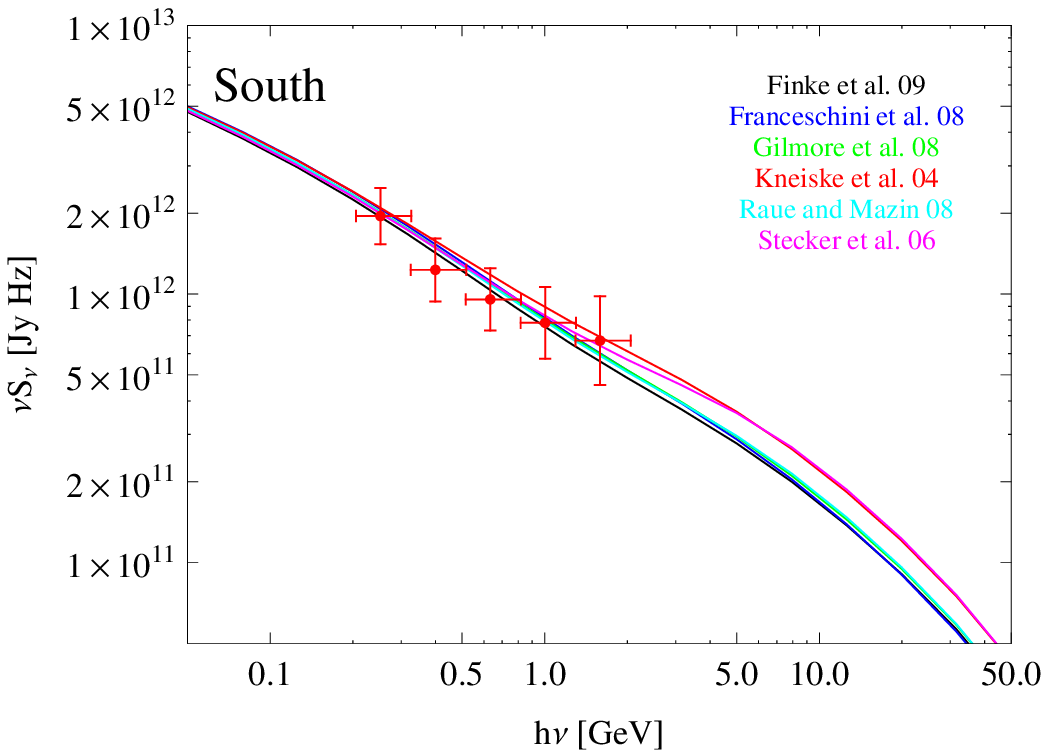}
  \end{center}
{\bf Fig.~S3.} Total IC fluxes of the northern [left] and southern 
[right] giant lobes computed for different EBL levels/spectral shapes, 
from \cite{fin09s} (black lines), \cite{fra08s} (blue), \cite{gil08s} 
(green), \cite{kne04s} (red), \cite{rau08s} (cyan), and \cite{ste06s} 
(magenta). \end{figure}

\begin{table}  
\begin{center}
\begin{tabular}{lcccc}
\hline\hline
\multicolumn{1}{l}{} & \multicolumn{2}{c}{North} & \multicolumn{2}{c}{South} \\
\hline
\multicolumn{1}{l}{Region}       &\multicolumn{1}{|c}{1} &\multicolumn{1}{|c}{2} &\multicolumn{1}{|c}{4} &\multicolumn{1}{|c}{5}\\
\hline
\multicolumn{1}{l}{Distance from core [kpc]} &\multicolumn{1}{|c}{226} &\multicolumn{1}{|c}{125} &\multicolumn{1}{|c}{131} &\multicolumn{1}{|c}{217}\\
\multicolumn{1}{l}{Cylindrical dimensions ($L,r$) [kpc]} &\multicolumn{1}{|c}{($122,81$)}  
 &\multicolumn{1}{|c}{($73,70.5$)}  &\multicolumn{1}{|c}{($86,78$)}  &\multicolumn{1}{|c}{($87, 69.5$)}\\
\multicolumn{1}{l}{$F$($>$100 MeV) [$10^{-7}$ ph cm$^{-2}$ s$^{-1}$]}    &\multicolumn{2}{|c}{0.77 (+0.23/--0.19)}                &\multicolumn{2}{|c}{1.09 (+0.24/--0.21)} \\
\multicolumn{1}{l}{Photon index}       &\multicolumn{2}{|c}{2.52 (+0.16/--0.19)}          &\multicolumn{2}{|c}{2.60 (+0.14/--0.15)} \\
\hline
\multicolumn{1}{l}{$k_{\rm e}$ [10$^{-10}$ cm$^{-3}$]} &\multicolumn{1}{|c}{86} &\multicolumn{1}{|c}{161} &\multicolumn{1}{|c}{0.016} &\multicolumn{1}{|c}{1.7} \\
\multicolumn{1}{l}{$s_{\rm 1}$}        &\multicolumn{2}{|c}{2.1}                       &\multicolumn{1}{|c}{1.1} &\multicolumn{1}{|c}{1.6} \\
\multicolumn{1}{l}{$s_{\rm 2}$}        &\multicolumn{2}{|c}{3.0}                       &\multicolumn{1}{|c}{3.4} &\multicolumn{1}{|c}{3.7} \\
\multicolumn{1}{l}{$\gamma_{\rm min}$} &\multicolumn{2}{|c}{1}       &\multicolumn{2}{|c}{1} \\
\multicolumn{1}{l}{$\gamma_{\rm br}$}  &\multicolumn{2}{|c}{$3.6 \times 10^{4}$}       &\multicolumn{2}{|c}{$3.6 \times 10^{4}$} \\
\multicolumn{1}{l}{$\gamma_{\rm max}$} &\multicolumn{2}{|c}{$3.3 \times 10^{5}$}       &\multicolumn{2}{|c}{$3.3 \times 10^{5}$} \\
\multicolumn{1}{l}{$B$ [$\mu$G]}       &\multicolumn{2}{|c}{0.89}                      &\multicolumn{2}{|c}{0.85} \\
\multicolumn{1}{l}{$U_{\rm e}/U_{\rm B}$}       &\multicolumn{2}{|c}{4.3}                      &\multicolumn{2}{|c}{1.8} \\
\multicolumn{1}{l}{$p$ [10$^{-14}$ erg cm$^{-3}$]}       &\multicolumn{2}{|c}{5.6}                      &\multicolumn{2}{|c}{2.7} \\
\hline
\end{tabular}
\label{table-params}
\end{center}
{\bf Table~S1.} Measured \& model parameters for the Cen~A giant lobes.
The quoted distances are from the core to the centroid of the different 
regions. Errors in the flux and photon indices are statistical only -- 
see Table~S3 for a summary of the systematic errors. 
\end{table}

\begin{table}
\begin{center}
\begin{tabular}{lrr}
\hline\hline
Name & \multicolumn{1}{c}{$\alpha_{\rm J2000}$} & \multicolumn{1}{c}{$\delta_{\rm J2000}$} \\
\hline
1FGL~J1300.9$-$3745  & $13^{\rm h}00^{\rm m}54^{\rm s}$ & $-37^\circ45'36"$ \\
1FGL~J1304.0$-$4622  & $13^{\rm h}04^{\rm m}05^{\rm s}$ & $-46^\circ22'06"$ \\
1FGL~J1304.3$-$4352  & $13^{\rm h}04^{\rm m}21^{\rm s}$ & $-43^\circ52'07"$ \\
1FGL~J1305.4$-$4928  & $13^{\rm h}05^{\rm m}28^{\rm s}$ & $-49^\circ28'15"$ \\
1FGL~J1307.0$-$4030  & $13^{\rm h}07^{\rm m}06^{\rm s}$ & $-40^\circ30'37"$ \\
1FGL~J1307.6$-$4259  & $13^{\rm h}07^{\rm m}38^{\rm s}$ & $-42^\circ59'58"$ \\
1FGL~J1320.1$-$4007  & $13^{\rm h}20^{\rm m}10^{\rm s}$ & $-40^\circ07'36"$ \\
1FGL~J1328.2$-$4729  & $13^{\rm h}28^{\rm m}12^{\rm s}$ & $-47^\circ29'56"$ \\
1FGL~J1334.2$-$4448  & $13^{\rm h}34^{\rm m}15^{\rm s}$ & $-44^\circ48'48"$ \\
1FGL~J1347.8$-$3751  & $13^{\rm h}47^{\rm m}52^{\rm s}$ & $-37^\circ51'18"$ \\
1FGL~J1400.1$-$3743  & $14^{\rm h}00^{\rm m}08^{\rm s}$ & $-37^\circ43'04"$ \\
\hline
\end{tabular}
\label{table-ptsrc}
\end{center}
{\bf Table~S2.} Positions of point sources detected by the $Fermi$-LAT within
the ROI that have been included in the background model.
\end{table}

\begin{table}
\begin{center}
\begin{tabular}{lcccccc}
\hline\hline
\multicolumn{1}{l}{Parameter} & 
\multicolumn{1}{c}{Value} & 
\multicolumn{1}{c}{Stat.} & 
\multicolumn{1}{c}{Sys.\ ($A_{\rm eff}$)} & 
\multicolumn{1}{c}{Sys.\ (diff.)} & 
\multicolumn{1}{c}{Sys.\ (radius)} & 
\multicolumn{1}{c}{Sys.\ (total)} \\
\hline
Core flux 	 & 1.50 & +0.25/--0.22 & +0.12/--0.11 & +0.02/--0.03 & +0.35/--0.00 & $\pm$0.37 \\
Core index 	 & 2.67 & +0.10/--0.10 & +0.06/--0.06 & +0.02/--0.00 & +0.00/--0.05 & $\pm$0.08 \\
North lobe flux  & 0.77 & +0.23/--0.19 & +0.23/--0.16 & +0.30/--0.14 & +0.00/--0.09 & $\pm$0.39 \\
North lobe index & 2.52 & +0.16/--0.19 & +0.18/--0.20 & +0.11/--0.14 & +0.06/--0.00 & $\pm$0.25 \\
South lobe flux  & 1.09 & +0.24/--0.21 & +0.25/--0.21 & +0.14/--0.16 & +0.00/--0.11 & $\pm$0.32 \\
South lobe index & 2.60 & +0.14/--0.15 & +0.17/--0.15 & +0.10/--0.05 & +0.00/--0.01 & $\pm$0.20 \\
\hline
\end{tabular}
\label{table-error}
\end{center}
{\bf Table~S3.} Summary of statistical and systematic errors.
The fluxes [$10^{-7}$ ph cm$^{-2}$ s$^{-1}$] quoted are in the $>100$ MeV band. In addition to the 
statistical (Stat.) errors, the sources of systematic (Sys.) 
uncertainties are due to limited knowledge about the LAT effective area 
($A_{\rm eff}$), variations of the diffuse model components (diff.), and core 
cut-out radius. Quoted are the minimum/maximum parameter differences 
from the values derived. The maximum values of each source of systematic error are added 
in quadrature to give the total systematic error. 
\end{table}

\end{document}